# Microstrip Line Based Complementary Resonant Structure For Dielectric Characterization

**Subhadip Roy, Pronoy Das, Puspita Parui, Anuvab Nandi and Chiranjib Mitra***

Department of Physical Sciences, Indian Institute of Science Education and Research, Kolkata, India

*Corresponding author: chiranjib@iiserkol.ac.in*

**Abstract**

In this work, a complementary resonant structure etched on the ground plane of a microstrip line is proposed for characterizing dielectric materials. The resonant sensor is designed to operate in S-band (2 – 4 GHz). The sensor is designed in an electromagnetic simulator to generate its transmission response, electric and magnetic field maps. A numerical model of the sensor is established to extract the electric permittivity of dielectric samples. The sensor is fabricated on a soft microwave laminate using a rapid photolithography technique. The electric permittivity values of wood, Teflon, and RT/duroid® 5880LZ are determined by using the sensor. The permittivity values are found consistent with those available in the literature.

## Introduction

Dielectric permittivity ($\varepsilon_r$) is an essential physical parameter which along with permeability, characterizes the interaction of a material with electromagnetic radiation [1]. Accurate characterization of $\varepsilon_r$ is of utmost importance for adequately modeling and designing microwave devices [2].

The resonance perturbation method is one of the widely used methods for determining the value of $\varepsilon_r$ [3]. In this method, the resonant characteristics of a sensor get modified by the material under test (MUT). This change yields information on the MUT's $\varepsilon_r$ value.

In the following sections, details about the sensor's geometry will be given. A comparison between the simulated and measured transmission characteristics of the device will be made. The generated electric and magnetic surface field maps of the device are shown. The sensor's numerical modeling [4] incorporates two correction terms that compensate for fabrication limitation and sample preparation. The $\varepsilon_r$ values of wood, Teflon, and RT/duroid® 5880LZ are determined here and are found to be consistent with literature.

## Sensor designing

The developed sensor is based on a microstrip line variant of the planar transmission line technology. The dielectric sensing region is a defected ground structure in the form of a conjoined meander and interdigital capacitor. The sensor is excited by the signal trace on the other side of the dielectric substrate. The geometry of the sensor is shown in fig. 1(a). The width of the signal trace is 1.2 mm. Other dimensions of the device are shown in fig. 1(b).

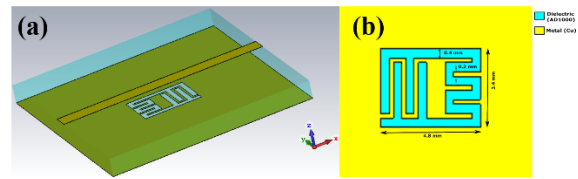

**Fig. 1.** **(a)** Geometry of the dielectric sensor. **(b)** Dimensions of the sensor region on the ground plane.

The sensor is fabricated on double-sided copper-clad microwave laminate Rogers AD1000 [5] using a rapid photolithography technique [6], [7], [8]. The fabricated sensor is shown in figs. 2 (a) and 2 (b).

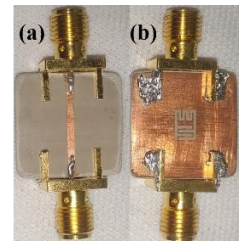

**Fig. 2.** The designed sensor. **(a)** Signal trace **(b)** Sensing region.

The electromagnetic field map of the sensor, as depicted in figs. 3 (a) and 3 (b) shows that the meander region has an electric field maxima and a magnetic field minima. Hence the MUTs are placed in this region for dielectric measurement.

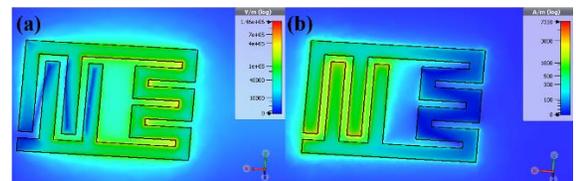

**Fig. 3.** **(a)** Electric field surface map. **(b)** Magnetic field surface map.

The transmission response of the device is measured using an R&S ZVA-24 vector network analyzer. The comparison between simulated and measured transmission responses are shown in fig. 4.

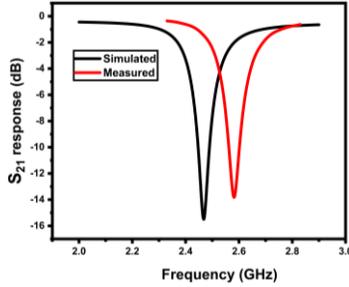

**Fig. 4.** Simulated and measured transmission response. The measured unloaded resonant frequency is 2.5815 GHz, whereas the simulated unloaded resonant frequency is 2.467 GHz.

## Measurement Procedure

The unloaded sensor can be modeled as an LCR circuit with resonant frequency $f$ given as

$$f = \frac{1}{2\pi\sqrt{LC}} \quad (1)$$

where $L$ and $C$ are the equivalent inductance and capacitance, respectively. When the sample is loaded on the meander region of the sensor, the equivalent capacitance changes, leading to a shift in $f$. It can be shown that the change obeys $f^{-2} \propto \varepsilon_r$ relation [4]. The numerical modeling involves designing square samples of length 3 mm with $\varepsilon_r$ varying from 1 to 10 and placing them on the meander region of the sensor in the simulation. The corresponding shift in $f$ is simulated and is used to generate a $f^{-2}$ vs $\varepsilon_r$ graph as shown in figs. 5(a) and (b). A linear fit of the graph gives the following relation.

$$f^{-2} = 0.00643\varepsilon_r + 0.15885 \quad (2)$$

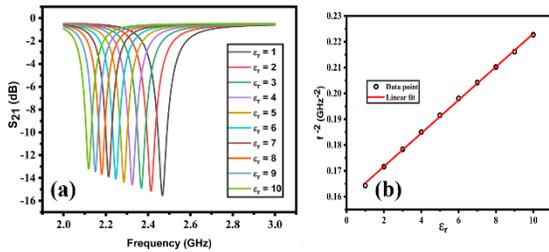

**Fig. 5. (a)** Variation of $f$ with $\varepsilon_r$ **(b)** $f^{-2}$ vs $\varepsilon_r$ graph along with the linear fit (red line).

Two correction terms described in [4] that correct the fabrication limitation and sample preparation are introduced in equation 2 to obtain the final empirical relation between $\varepsilon_r$ and measured $f$.

$$\varepsilon_r = \frac{(f - 0.1135 - 0.007243f)^{-2} - 0.15885}{0.00643} \quad (3)$$

The MUTs are loaded on the sensor, and $f$ is measured. The measured $f$ is used in equation 3 to obtain the final measured $\varepsilon_r$ values of the MUTs.

## Result and Discussion

The measured $\varepsilon_r$ values of the MUTs are listed in table 1. They have been compared with those available in the literature, and the corresponding error percentages have been calculated.

**Table 1.** Comparison between measured and reference $\varepsilon_r$ values of the MUTs

| Sample | Measured $f$ (GHz) | Reference $\varepsilon_r$ | Measured $\varepsilon_r$ | % error |
|---|---|---|---|---|
| 5880LZ | 2.548 | 2.0 [9] | 1.92 | ~ 4 |
| Teflon | 2.541 | 2.1 [4] | 2.07 | ~ 1.4 |
| Wood | 2.5285 | 2.3 [1] | 2.35 | ~ 2.2 |

From the table, it can be seen that the maximum error percentage in determining $\varepsilon_r$ values is close to 4%. Hence it can be concluded that the sensor extracts the permittivity value of MUTs with considerable accuracy. The sensor can be further improved and extended to simultaneous permittivity measurement.

## Acknowledgment

The authors acknowledge the Ministry of Education (MoE) for funding this work. S.R. and P.P. thank the Council of Scientific and Industrial Research (CSIR) for their fellowship. The authors are grateful to Prof Bhaskar Gupta, Department of Electronics & Telecommunication Engineering, Jadavpur University, for simulation facilities. The authors thank Roger Corporation, USA, for the microwave laminate samples.

## References

1. X. Lu *et al.*, *J. Phys. D*. **53,** 9 (2020).
2. M. Saadat-Safa, V. Nayyeri, M. Khanjarian, M. Soleimani, O. M. Ramahi, *IEEE Trans. Microw. Theory Techn*. **67**, 806–814 (2019).
3. K. T. Muhammed Shafi, A. K. Jha, M. J. Akhtar, *IEEE Sens. J*. **17**, 5479–5486 (2017).
4. A. Kapoor, P. K. Varshney, M. J. Akhtar, *Microw. Opt. Technol. Lett*. **62**, 2835–2840 (2020).
5. Rogers Corporation AD1000 datasheet.
6. S. Roy, S. Saha, J. Sarkar, C. Mitra, *Eur. Phys. J. Appl. Phys*. **90** (2020),
7. S. Roy, A. Nandi, P. Das, C. Mitra, *IOP SciNotes* **1**, 035202 (2020).
8. S. Roy, A. Nandi, P. Das, C. Mitra, *J. Phys. D: Appl. Phys*. **54,** (2021).
9. Rogers Corporation RT/duroid® 5880LZ datasheet.